\documentclass[conference]{IEEEtran}

\usepackage[hidelinks]{hyperref}

\usepackage[utf8]{inputenc}
\usepackage[english]{babel}

\usepackage{bm}
\usepackage{amsthm}
\usepackage{comment}
\usepackage{xpatch}
\usepackage{color}
\usepackage{stfloats}
\usepackage[abs]{overpic}
\usepackage{cite}
\usepackage{amssymb}
\usepackage{amsmath}
\usepackage{graphicx}
\usepackage{algorithm}
\usepackage{algorithmic}
\usepackage{url} 

\usepackage[figurename=Fig.]{caption}
\usepackage{comment}

\usepackage{enumerate}

\renewcommand\IEEEkeywordsname{Index Terms}

\newcommand{\numSAPs}{L}
\newcommand{\numAntennasSAP}{N}
\newcommand{\numSUEs}{K_{\rm s}}

\newcommand{\numAntennasPBS}{M}
\newcommand{\numPUEs}{K_{\rm p}}

\newcommand{\hyphen}{\text{-}}
\newcommand{\taup}{\tau_{\rm p}}
\newcommand{\tauc}{\tau_{\rm c}}

\newcommand{\pilot}{\boldsymbol{\phi}}
\newcommand{\sspilotPow}{\eta_{\rm s}}
\newcommand{\pppilotPow}{\eta_{\rm p}}

\newcommand{\sschannel}[1]{\mathbf{h}_{{#1}}}
\newcommand{\spchannel}[1]{\mathbf{u}_{{\rm sp}\hyphen#1}}
\newcommand{\pschannel}[1]{\mathbf{u}_{{\rm ps}\hyphen#1}}
\newcommand{\ppchannel}[1]{\mathbf{g}_{{#1}}}

\newcommand{\sschanCov}{\mathbf{R}}
\newcommand{\sschanEstCov}{\widehat{\mathbf{R}}}
\newcommand{\sschanErrCov}{\widetilde{\mathbf{R}}}
\newcommand{\ssYpCorr}[2]{\boldsymbol{\Psi}_{t_{#1}#2}}

\newcommand{\spchanCov}[1]{\mathbf{C}_{{\rm sp}\hyphen#1}}
\newcommand{\pschanCov}[1]{\mathbf{C}_{{\rm ps}\hyphen#1}}
\newcommand{\ppchanCov}[1]{\mathbf{D}_{#1}}

\newcommand{\sschannelest}{\widehat{\mathbf{h}}}
\newcommand{\sschannelerr}{\widetilde{\mathbf{h}}}

\newcommand{\ppchannelest}[1]{\widehat{\mathbf{g}}_{{#1}}}

\newcommand{\mrprecoder}[2]{\frac{\widehat{\mathbf{h}}_{#1#2}}{\sqrt{\mathbb{E}\{\Vert \widehat{\mathbf{h}}_{#1#2}\Vert^2\}}}}
\newcommand{\sschanprecoder}{\mathbf{w}}

\newcommand{\Pmax}{P_{\rm max}}

\newcommand{\Ithn}{\mathcal{I}_{\rm th}/\sigma^2}
\newcommand{\Ith}{\mathcal{I}_{\rm th}}

\newcommand{\norm}[1]{\left\Vert#1\right\Vert}
\newcommand{\absL}[1]{\left\vert#1\right\vert}

\newcommand{\complexm}[2]{\mathbb{C}^{#1\times #2}}

\newcommand{\CN}[2]{\mathcal{CN}{(#1,#2)}}
\newcommand{\expect}[1]{\mathbb{E}\{#1\}}

\newcommand{\trace}[1]{{\rm Tr}(#1)}

\IEEEoverridecommandlockouts
\begin{document}
%
\title{Energy-Efficient Power Allocation for an Underlay Spectrum Sharing RadioWeaves Network}


%
\author{Zakir Hussain Shaik$^{*}$, Rimalapudi Sarvendranath$^{\dagger}$ and Erik G. Larsson$^{*}$ \\
	$^{*}$Department of Electrical Engineering (ISY), Link{\"o}ping University, Link{\"o}ping, Sweden\\$^{\dagger}$Department of Electronics and Electrical Engineering, Indian Institute of Technology Guwahati, Guwahati, India\\
	Email: zakir.hussain.shaik@liu.se, sarvendranath@iitg.ac.in, erik.g.larsson@liu.se\thanks{At the time the paper was submitted,  Sarvendranath was with the Division of Communications Systems, Department of Electrical Engineering, Link{\"o}ping University 58183, Link{\"o}ping, Sweden. He is now with the Indian Institute of Technology Guwahati, Guwahati-781039, India.\newline This work was partially funded by the REINDEER project of the European Union‘s Horizon 2020 research and innovation program under grant agreement No. 101013425. This work was also partially supported by ELLIIT and KAW.}}


\maketitle

\begin{abstract}
RadioWeaves network operates a large number of distributed antennas using cell-free architecture to provide high data rates and support a large number of users. Operating this network in an energy-efficient manner in the limited available spectrum is crucial. Therefore, we consider energy efficiency (EE) maximization of a RadioWeaves network that shares spectrum with a collocated primary network in underlay mode. 
To simplify the problem, we lower bound the non-convex EE objective function to form a convex problem. We then propose a downlink power allocation policy that maximizes the EE of the secondary RadioWeaves network subject to power constraint at each access point and interference constraint at each primary user. 
Our numerical results investigate the secondary system's performance in interference, power, and EE constrained regimes with correlated fading channels. Furthermore, they show that the proposed power allocation scheme performs significantly better than the simpler equal power allocation scheme.
\end{abstract}

%
\IEEEpeerreviewmaketitle

\vspace{-1.2mm}
\section{Introduction}

RadioWeaves is an emerging technology build upon the concepts of cell-free massive multiple-input-multiple-output (CF-mMIMO) and possibly large intelligent surfaces wherein the antennas and underlying signal processing circuitry are weaved into large surface areas such as conventional buildings and objects \cite{VanRadioWeaves,UnniRadioWeaves}.
The distributed infrastructure of RadioWeaves with many antennas provides favorable path loss conditions and also leverages the benefits of CF-mMIMO such as macro-diversity i.e., robustness against signal blockage as user-equipments (UEs) are highly likely to be close to some of the antennas \cite{Hien,interdonato2019ubiquitous}. 
Most of the wireless spectrum in lower frequencies, which have good propagation characteristics, is already allocated, thus {\em spectrum sharing} is crucial to enable RadioWeaves.
Spectrum sharing helps RadioWeaves to support high data rates and large number of devices while utilizing the scarce spectrum efficiently~\cite{amarsuryaTCoM,Rezaei2,Rezaei}. 
 
Accepting the need for spectrum sharing, even the spectrum regulators such as Federal Communications Commission opened $3~{\rm GHz}$ ($3.55 - 3.70~{\rm GHz}$) and $6~{\rm GHz}$ bands ($5.925 - 7.125~{\rm GHz}$) for shared operations \cite{fcc_report3,fcc_report6}.
With indoor operations as one of its main focus, RadioWeaves, can take advantage of $350~{\rm MHz}$ of bandwidth in the $6~{\rm GHz}$ band that is open for shared indoor operations.  
Spectrum sharing has also gained practical attention and is part of the current and next-generation wireless standards such as long term evolution (LTE)-license assisted access, MulteFire, Citizen's broadband radio service, 5G new radio unlicensed, and IEEE $802.11{\rm be}$~\cite{au2019ieee}.
Among different modes of spectrum sharing, 
in underlay mode, a low-priority secondary network (SN) transmits concurrently with an incumbent primary network (PN)~\cite{TanabSurvey, amarsuryaTCoM}. 
Hence, the SN must ensure that the interference caused to the PN is below permissible levels. 

{\em Spectrum Sharing}:
In~\cite{amarsuryaTCoM}, authors studied a multi-objective power allocation policy to achieve max-min fairness with common minimum signal-to-interference-plus-noise-ratio (SINR) for both PN and SN. This was done for orthogonal multiple access (OMA) and non-orthogonal multiple access (NOMA) systems. In \cite{Rezaei2}, authors investigated the sum-rate maximization of CF-mMIMO system operating as SN and massive multiple-input-multiple-output (mMIMO) system as PN. On similar lines, \cite{Rezaei} studied sum-rate maximization of a CF-mMIMO NOMA system operating as SN and mMIMO as PN.

{\em Energy Efficiency (EE)}, which is defined as the number of information bits that can be reliably transmitted per unit energy, is an important metric of performance for a communication network.
Maximizing EE  provides trade-off between date rate and the energy spent. 
The EE of a CF-mMIMO was studied in \cite{Zappone,HienEE,Tan}. In \cite{Zappone}, the authors proposed a low-complexity iterative algorithm to maximize the EE of a CF-mMIMO system.
In \cite{HienEE}, the authors considered a CF-mMIMO system with uncorrelated channels and MR precoding. They proposed a downlink power allocation policy to maximize the EE using second order conic programming. 
In~\cite{Tan}, EE based resource allocation was explored for a layered-division multiplexing based non-orthogonal multicast and unicast transmission systems with simultaneous wireless information and power transfer. EE has been studied extensively for networks other than CF-mMIMO for instance \cite{sboui2015energy}.
EE maximization is also crucial for a distributed network such as RadioWeaves operating in underlay spectrum sharing mode. It enables efficient utilization of the energy and spectrum.
To the best of our knowledge there is no prior work on maximizing EE of a RadioWeave network operating in underlay spectrum sharing mode. 

{\em Focus and Contributions:}
We focus on a downlink underlay centralized RadioWeave network with multiple APs connected to a CPU and serve multiple secondary UEs. 
It share spectrum with a collocated PN with multiple users. 
Our model is novel and practical in the following aspects: ({\it i}) Firstly, the EE problem is not studied for an underlay RadioWeave network.
({\it ii}) Secondly, we assume imperfect CSI at the APs to perform precoding, which is more practical assumption compared to widely considered perfect CSI. Furthermore, we consider correlated channel gains at each AP.
({\it iii}) Thirdly, only channel statistics, which vary slowly compared to the instantaneous channel gains, are assumed at the CPU. 
({\it iv}) Lastly, we consider practically motivated average interference constraint~\cite{amarsuryaTCoM,Sarvendranath_2014_TCOM} and maximum transmit power constraints, which considers limitations of the power amplifiers.

Specific contributions are as follows:
\begin{itemize}
\item First, we reformulate the non-convex EE optimization problem into a convex problem by lower bounding the objective function. For this modified problem, we propose a downlink power allocation algorithm that attains the optimal solution.

\item We provide a closed-form expression for the achievable SE of SN RadioWeave network with underlay spectrum sharing.

\item Our numerical results study the impact of interference constraints on the EE of a RadioWeave network. They also show that the proposed power allocation policy performs significantly better than the simpler policy that allocates equal power to each user.
\end{itemize}

\textit{Notations:} Boldface lowercase letters, $\mathbf{a}$, denote column vectors and boldface uppercase letters, $\mathbf{A}$, denote matrices. The superscripts $(\cdot)^*,~(\cdot)^T,$ and $(\cdot)^H$ denote conjugate, transpose, and Hermitian transpose, respectively. The $N\times N$ identity matrix is denoted by $\mathbf{I}_N$. The absolute value of a scalar and $l_2$ norm of a vector are denoted by $\vert \cdot \vert$, and $\Vert \cdot \Vert$, respectively. We denote expected value of a random variable  $x$ as $\expect{x}$. We use $\mathbf{z} \sim \mathcal{CN}\left(\mathbf{0},\mathbf{C}\right)$ to denote a circularly symmetric complex Gaussian random vector with covariance matrix $\mathbf{C}$. 

\section{System Model And Channel Estimation}\label{SystemModel}
\begin{figure}[!t]
\centering
\includegraphics[width=.7\linewidth]{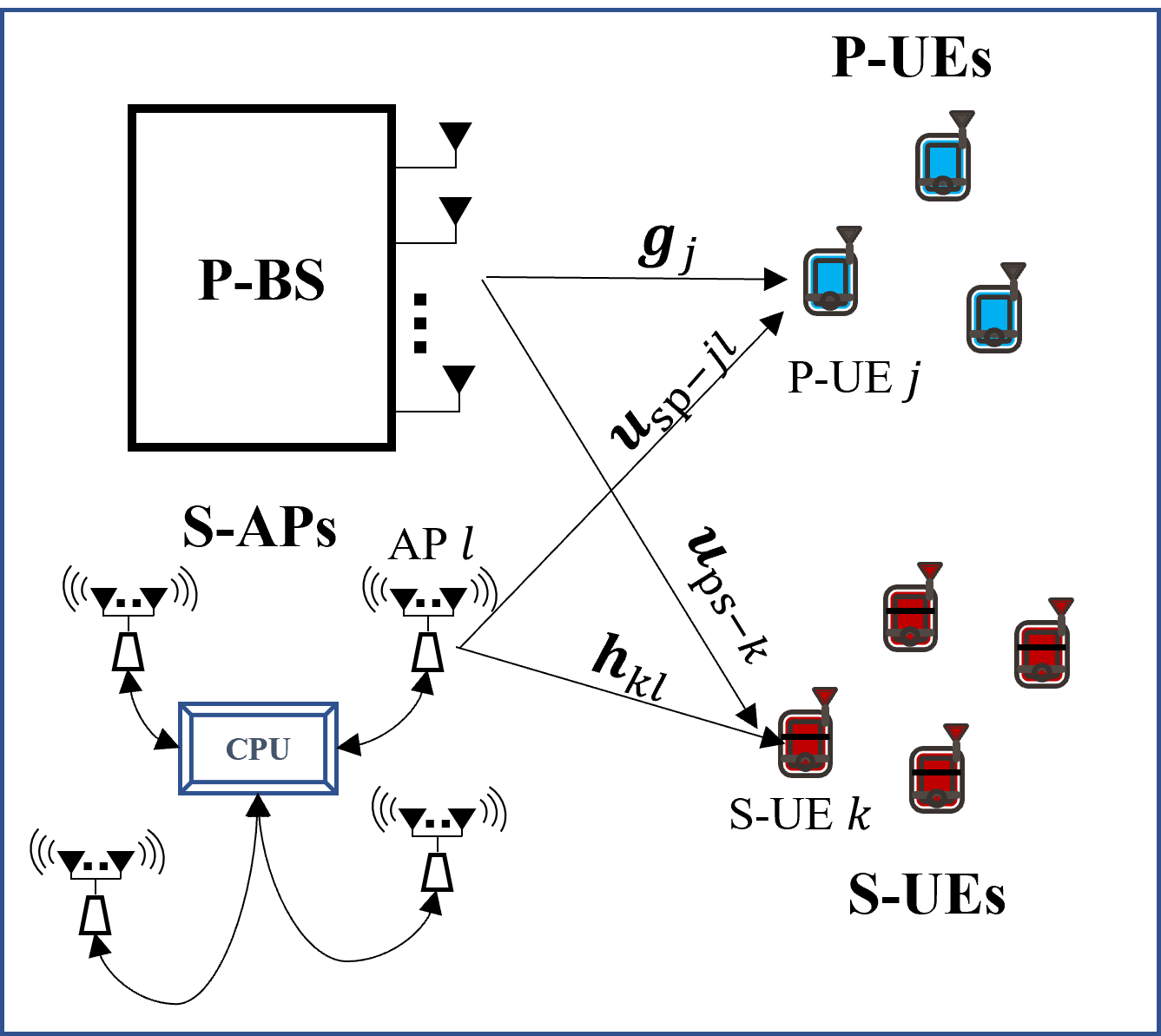}
\caption{Network Model.}
\label{fig:SysModel}
\end{figure}

The system model is shown in Fig.~\ref{fig:SysModel}.  In it, a RadioWeave SN shares spectrum with a PN in underlay mode~\cite{amarsuryaTCoM}.
The primary base station (P-BS) with $\numAntennasPBS$ antennas serves $\numPUEs$ single-antenna primary user equipments (P-UEs). 
The SN consists of  $\numSAPs$ secondary APs (S-APs), each equipped with $\numAntennasSAP$ antennas, and serves $\numSUEs$ single-antenna secondary user equipments (S-UEs). 
Both PN and SN operate in TDD mode. Furthermore, they operate synchronously performing uplink and downlink operations at the same time. 
We consider correlated Rayleigh fading for all the links. Let $\sschannel{kl}~\sim ~\CN{\mathbf{0}}{\sschanCov_{kl}}$,  $\spchannel{jl}~\sim ~\CN{\mathbf{0}}{\spchanCov{jl}}$, $\pschannel{i}~\sim ~\CN{\mathbf{0}}{\pschanCov{i}}$ and $\mathbf{g}_m~\sim ~\CN{\mathbf{0}}{\ppchanCov{m}}$ denote the channel gains from  $l\hyphen{\rm th}$ S-AP to $k\hyphen{\rm th}$ S-UE, $l\hyphen{\rm th}$ S-AP to $j\hyphen{\rm th}$ P-UE, from P-BS to $i\hyphen{\rm th}$ S-UE, and P-BS to $m\hyphen{\rm th}$ P-SU respectively, and  $\sschanCov_{kl}\in \complexm{N}{N}$, $\spchanCov{jl}\in \complexm{N}{N}$, $\pschanCov{i}\in \complexm{M}{M}$ and $\ppchanCov{m}\in \complexm{M}{M}$ denote corresponding channels correlation matrices. We assume that the channel gains $\sschannel{kl}$, $\spchannel{jl}$,  $\pschannel{i}$ and $\ppchannel{i}$ are independent of each other. We assume that the channel statistics are known at the S-APs and the CPU. We now describe the uplink channel estimation and downlink data transfer phases of the SN.

\subsection{Channel Estimation}
Let $\tauc$ denote the length of the coherence interval over which channel is constant and $\taup \leq \tauc$ denote the length of the pilot sequence. 
We consider $\taup$ mutually orthogonal pilot sequences, $\boldsymbol{\Phi}= [\pilot_1,\ldots,\pilot_{\rm \taup}]\in \complexm{\taup}{\taup}$ such that $\norm{\pilot_t} = \sqrt{\taup}$, $t=1,\ldots,\taup$ are distributed between secondary and primary networks. The pilot allocation scheme is based on the categorization of pilot sequences as $\boldsymbol{\Phi} = [\boldsymbol{\Phi}_0,\boldsymbol{\Phi}_{\rm p},\boldsymbol{\Phi}_{\rm s}]$ where $\boldsymbol{\Phi}_0 \in \complexm{\taup}{\tau_1}$ denote the pilot sequences shared by both PN and SN, $\boldsymbol{\Phi}_{\rm p} \in \complexm{\taup}{\tau_2}$ denote the pilot sequences allocated to only PN and $\boldsymbol{\Phi}_{\rm s} \in \complexm{\taup}{\tau_3}$ denote the pilot sequences allocated to only SN, with $\ \tau_1 + \tau_2 + \tau_3 = \taup$.

S-APs employ minimum-mean square estimation (MMSE) to obtain channel estimates $\sschannelest_{kl}$. Let $t_k\in \{1,\ldots,\tau_{\rm p}\}$ denote the  index of the pilot sequence assigned to S-UE $k$. 
Furthermore, let $\mathcal{S}_{k}^{\rm p}$ and $\mathcal{S}_{k}^{\rm s}$ be the set of P-UEs and S-UEs, respectively, that share the same pilot sequence as S-UE $k$. 
The pilot signal received at S-AP $l$ is given by
\begin{equation}
\mathbf{Y}_l^{\rm p} =  \sqrt{\sspilotPow}\sum_{i=1}^{\numSUEs}\sschannel{il}\pilot_{t_i}^T + \sqrt{\pppilotPow}\sum_{j=1}^{\numPUEs}\spchannel{jl}\pilot_{t_j}^T + \mathbf{N}_l,
\end{equation}
where $\sspilotPow \geq 0$ and $\pppilotPow \geq 0$ are the pilot powers used by the S-UEs and P-UEs, respectively, and $\mathbf{N}_l\in \complexm{N}{\taup}$ is the noise at the receiver with independent and identical entries drawn from $\CN{0}{\sigma^2}$ with $\sigma^2$ being the noise power. 

The projection of $\mathbf{Y}_l^{\rm p}$ on $\pilot_{t_k}$ is given by
\begin{equation}
\begin{aligned}
	\mathbf{y}_{t_k l}^{\rm p} &= \mathbf{Y}_l^{\rm p}{\pilot_{t_k}^{*}}/{\sqrt{\taup}},\\
	& =\sqrt{\sspilotPow\taup}\sum_{i\in \mathcal{S}_{k}^{\rm s} } \sschannel{il} + \sqrt{\pppilotPow\taup}\sum_{j\in \mathcal{S}_{k}^{\rm p}} \spchannel{jl} + \mathbf{n}_{{t_k}l},
\end{aligned}
\end{equation}
where $\mathbf{n}_{{t_k}l} =  \mathbf{N}_l\pilot_{t_k}^{*}/\sqrt{\taup}\sim \CN{\mathbf{0}}{\sigma^2 \mathbf{I}_N}$  is noise.
The MMSE channel estimate $\sschannelest_{kl}$ based on $\mathbf{y}_{t_k l}^{\rm p}$ is
\begin{equation}\label{ssChanEst}
\sschannelest_{kl} = \sqrt{\sspilotPow\taup}\sschanCov_{kl}\ssYpCorr{k}{l}^{-1}\mathbf{y}_{t_k l}^{\rm p},
\end{equation}
where
\begin{equation}
\ssYpCorr{k}{l} =  \sspilotPow\taup\sum_{i\in \mathcal{S}_{k}^{\rm s} } \sschanCov_{il} + \pppilotPow\taup\sum_{j\in \mathcal{S}_{k}^{\rm p} }  \spchanCov{jl} + \sigma^2\mathbf{I}_{N}.
\end{equation}
 Also, from MMSE estimation theory, it follows that $\sschannelest_{kl}\sim\CN{\mathbf{0}}{\sschanEstCov_{kl}}$ and the estimation error $\sschannelerr_{kl} = \sschannel{kl}- \sschannelest_{kl}\sim\CN{\mathbf{0}}{\sschanErrCov_{kl}}$ where $\sschanEstCov_{kl} = \sspilotPow\taup\sschanCov_{kl}\ssYpCorr{k}{l}^{-1}\sschanCov_{kl}$ and $\sschanErrCov_{kl} = \sschanCov_{kl} - \sschanEstCov_{kl}$.

\subsection{Downlink Payload Transmission}

Let $p_{il}\geq 0$ denote the transmit power allocated for S-AP $l$ to transmit message symbol  $q_i$ with $\mathbb{E}\{\vert q_i \vert^2\} = 1$ to S-UE $i$. 
It employs MR precoding as a function of channel estimates, which is given by~\cite{OzlemCellFreeBook}

\begin{equation}\label{mrPrecod}
	\mathbf{w}_{il} = \mrprecoder{i}{l},
\end{equation}
and transmits
\begin{equation}
\mathbf{x}_l = \sum_{i=1}^{\numSUEs}\sqrt{p_{il}}\mathbf{w}_{il}q_i,\quad l = 1,\ldots,\numSAPs.
\end{equation}
Therefore, the  signal received at  S-UE $k$ is given by
\begin{align}
y_k & = \sum_{l=1}^\numSAPs \mathbf{h}_{kl}^H \mathbf{x}_l + d_k+ n_k,\label{recSigS-UE}\\
&= \sum_{l=1}^\numSAPs \sqrt{p_{kl}}\mathbf{h}_{kl}^H \mathbf{w}_{kl}q_k + \sum_{\substack{i=1 \\ i\neq k}}^{\numSUEs}\sum_{l=1}^{\numSAPs}\sqrt{p_{il}}\mathbf{h}_{kl}^H \mathbf{w}_{il}q_i + d_k+ n_k,\notag
\end{align}
where $ d_k = \mathbf{a}_k^H\bar{\mathbf{x}}_p$ is the interference at the $k$th S-UE from the P-BS, $\bar{\mathbf{x}}_p$ is the primary transmit signal with correlation matrix $\mathbf{Q}_p =\expect{\bar{\mathbf{x}}_p\bar{\mathbf{x}}_p^H}$, and $n_k \sim \CN{0}{\sigma^2}$ is the additive thermal noise at S-UE $k$.  We note that $\bar{\mathbf{x}}_p$ is independent of $\mathbf{a}_k$ and that the interference signal $d_k$ is independent of the channels and data signals of the SN.

We assume that S-UE $k$ knows only the statistics of the effective channel gain from each of the S-APs i.e., $\mathbb{E}\left\{\mathbf{h}_{kl}^H \mathbf{w}_{kl}\right\}$, for $l = 1,\ldots,\numSAPs$.
For S-UE $k$, let ${\rm DK}_k=\sum_{l=1}^\numSAPs \sqrt{p_{kl}}\mathbb{E}\left\{\mathbf{h}_{kl}^H \mathbf{w}_{kl}\right\}q_k $ denote the desired message symbol $q_k$  transmitted over known effective channel gains. 
Adding and subtracting ${\rm DK}_k$ to \eqref{recSigS-UE}, $y_k$ can be rewritten as
\begin{equation}\label{recSigS-UE2}
y_k = {\rm DK}_k +{\rm DU}_k +  {\rm UI}_{k} + d_k + n_k
\end{equation}
where,
\begin{align}
{\rm DU}_k & =  \sum_{l=1}^\numSAPs \sqrt{p_{kl}}\mathbf{h}_{kl}^H \mathbf{w}_{kl}q_k -  {\rm DK}_k,\label{DUk}\\
{\rm UI}_{k} & = \sum_{\substack{i=1 \\ i\neq k}}^{\numSUEs}\sum_{l=1}^{\numSAPs}\sqrt{p_{il}}\mathbf{h}_{kl}^H \mathbf{w}_{il}q_i. \label{UIk}
\end{align}
The above notations ${\rm DU}_k$, ${\rm UI}_k$ can be interpreted as desired symbol $q_k$ transmitted over unknown channel and inter-user interference from other S-UEs, respectively.
From~\eqref{recSigS-UE2}, the effective SINR $\Gamma_k$ can be written as 
{\small
	\begin{equation}\label{sinr_k}
		\Gamma_k = \frac{\vert {\rm DK}_k \vert^2}{ \mathbb{E}\{\vert{\rm DU}_k \vert^2\} +  \mathbb{E}\{\vert {\rm UI}_{k} \vert^2\} + \varsigma_k^2},
	\end{equation}
} 
where $\varsigma_k^2 = \trace{\mathbf{Q}_p\pschanCov{k}} + \sigma^2$ represents interference-plus-noise-power at S-UE $k$.
Then, $\mathrm{SE}_k$, the achievable net SE for UE $k$ is given by
\begin{equation}\label{specEff_1}
\mathrm{SE}_k = \left(1-\frac{\tau_p}{\tau_c}\right)\mathrm{log}_2\left( 1 + \Gamma_k\right)  \ \rm{bit/s/Hz}.
\end{equation}
In literature the capacity bound presented in \eqref{specEff_1} is called as use-and-forget bound \cite{marzetta2016fundamentals}.

The SINR $\Gamma_k$ can be computed in closed form
\begin{equation}\label{sinr2}
\Gamma_k = \frac{\displaystyle\left(\sum_{l=1}^{L}\sqrt{p_{kl}}a_{kkl}\right)^2}{\displaystyle\sum_{i=1}^{\numSUEs}\sum_{l=1}^{\numSAPs} p_{il}b_{ikl} + \sum_{i=1,i\neq k}^{\numSUEs}\left(\sum_{l=1}^{L}\sqrt{p_{il}}a_{ikl}\right)^2+\sigma_k^2}
\end{equation}
where 
\begin{subequations}
	\begin{align}
	b_{ikl} &= \frac{\trace{\sschanCov_{il}\ssYpCorr{i}{l}^{-1}\sschanCov_{il}\sschanCov_{kl}}}{\trace{\sschanCov_{il}\ssYpCorr{i}{l}^{-1}\sschanCov_{il}}}\\
	a_{ikl} &= \mathbb{I}_{\mathcal{S}_{k}^{\rm s}}(i) \frac{\sspilotPow\taup\trace{\sschanCov_{il}\ssYpCorr{i}{l}^{-1}\sschanCov_{kl}}}{\sqrt{\sspilotPow\taup\sschanCov_{il}\ssYpCorr{i}{l}^{-1}\sschanCov_{il}}},
	\end{align}
\end{subequations}
with $\mathbb{I}_{\mathcal{S}_{k}^{\rm s}}(i) = 1$, when $i\in \mathcal{S}_k^{\rm s}$ and  $\mathbb{I}_{\mathcal{S}_{k}^{\rm s}}(i) = 0$, when $i\notin \mathcal{S}_k^{\rm s}$. 
Observe that if there is no pilot contamination then $a_{ikl} = 0, i\neq k$. For derivation of these terms we refer to \cite{OzlemCellFreeBook}.

\section{Problem Formulation}
Let $\mathbf{p}=[p_{11},\ldots,p_{K1},\ldots,p_{1L},\ldots,p_{KL}]^T$ denote the transmit power allocation vector. 
The total power consumed $P_T(\mathbf{p})$ at all APs during downlink transmission is given by
\begin{equation}
P_T(\mathbf{p}) = \zeta\sum_{l=1}^{\numSAPs}\sum_{i=1}^{\numSUEs}p_{il} + \xi\sum_{i=1}^{\numSUEs}B\cdot{\rm SE}_i(\mathbf{p}) + {\rm PC} ,
\end{equation}
where  $\zeta$ is the inverse of power amplifier efficiency, $B$ is the secondary system bandwidth,  $\xi$ is the throughput dependent front-haul power consumption factor expressed in Watt/bit/s, and ${\rm PC}$ is the total circuit power consumption.
The EE of the secondary system as a function of the transmit power allocation vector $\mathbf{p}$ is equal to the ratio between the sum throughput and the total power consumed at all S-APs, i.e.,

\begin{equation}\label{eeObj}
{\rm EE}(\mathbf{p}) = \frac{\sum_{k=1}^{\numSUEs}B\cdot{\rm SE}_k(\mathbf{p})}{P_{T}(\mathbf{p})}.
\end{equation}
An important point to note here is that the power allocation policy that maximizes~\eqref{eeObj} remains same when the throughput dependent power term, $P_T(\mathbf{p})$, is  replaced by a throughput independent power term $\overline{P}_T(\mathbf{p}) = \zeta\sum_{l=1}^{\numSAPs}\sum_{i=1}^{\numSUEs}p_{il} + {\rm PC}$. A proof for this observation follows similar steps as described in Appendix~B of~\cite{HienEE}. 
This change simplifies the problem at hand. 
Therefore, we use $\overline{P}_T(\mathbf{p})$ in the denominator of our objective function when we state our optimization problem.
 
We consider the following two constraints:
\begin{itemize}
	\item {\em Peak Transmit Power constraint at each S-AP}:
		The total transmit power at each S-AP is constrained by a maximum transmit power $P_{\textrm{max}}$, i.e., 
	\begin{equation} \label{consPatAP}
	\sum_{i=1}^{\numSUEs}p_{il} \leq P_{\rm{max}},\ l = 1,\cdots,\numSAPs.
	\end{equation}
	\item {\em Average Interference Constraint~\cite{amarsuryaTCoM,Sarvendranath_2014_TCOM}}: 
	The interference signal at P-UE $m$ is given by
	\begin{equation} \label{PUErx}
	y_m^{\rm{sp}} = \sum_{l=1}^\numSAPs \mathbf{u}_{ml}^H \mathbf{x}_l.
	\end{equation} 
	
Therefore, the average interference seen by the $m\hyphen{\rm th}$ P-UE is given by  
\begin{equation} \label{consIntatPUE}
	\begin{aligned}
		\mathbb{E}\{\vert \mathbf{y}_m^{\rm{sp}}\vert^2\} = \sum_{l=1}^{\numSAPs}\sum_{i=1}^{\numSUEs}p_{il}\vartheta_{iml} ,
	\end{aligned}
\end{equation}
where $\vartheta_{iml}=\trace{\sschanEstCov_{il}\spchanCov{ml}}/\trace{\sschanEstCov_{il}}$.
The average interference constraint limits the average interference at each P-UE to be below a threshold $\mathcal{I}_{\rm th}$ i.e., $\mathbb{E}\{\vert \mathbf{y}_m^{\rm{sp}}\vert^2\}\leq \mathcal{I}_{\rm th}$.
\end{itemize}

Our aim is to develop an optimal downlink power allocation policy, $\mathbf{p}^*$, that maximizes the EE of the secondary network subject to a peak transmit power constraint and the average interference constraint. It can be written as the following optimization problem: 
\begin{subequations} \label{prblm2}
\begin{align}
\underset{\mathbf{p}}{\rm{max.}}\quad & \dfrac{\sum_{k=1}^{\numSUEs}B\cdot{\rm SE}_k(\mathbf{p})}{\overline{P}_{T}(\mathbf{p})}, \label{eeObj2}\\
\text{s.t.} & \displaystyle\sum_{i=1}^{\numSUEs}p_{il} \leq P_{\rm{max}},\ l = 1,\ldots,\numSAPs,\label{conPwrAPs} \\
&\displaystyle\sum_{l=1}^{\numSAPs}\sum_{i=1}^{\numSUEs}p_{il}\vartheta_{iml} \leq  \mathcal{I}_{\rm th},\ m = 1,\ldots,\numPUEs\label{conInt}\\
&\displaystyle p_{il} \geq 0,\ i=1,\ldots,\numSUEs;\ l = 1,\ldots,\numSAPs.\label{conPstvPwr}
\end{align}
\end{subequations}
Though the constraints \eqref{conPwrAPs}, \eqref{conInt} and \eqref{conInt} are convex  the problem in \eqref{prblm2} is non-convex as the objective function \eqref{eeObj2} is non-convex. To circumvent the non-convexity problem, we provide a suitable convex lower bound on EE problem \eqref{prblm2} and solve it instead. Before we state the convex lower bound problem, we recollect the following two facts: $(i)$ the geometric mean of positive variables is concave and $(ii)$ the trace of product of any two positive semi-definite matrices is non-negative. Thus, all the terms in \eqref{sinr2} are non-negative. Moreover, the terms in the numerator and denominator of SINR $\Gamma_k$ are positive weighted sums of geometric mean terms which are concave.

Although, the numerator and denominator terms of $\Gamma_k$ are concave, $\Gamma_k$ is not concave. So to tackle this, we rewrite the sum SE as the difference of two concave functions (the logarithm of a concave function is concave) i.e., $\sum_{k=1}^{K}{\rm SE}_k = f_{1}(\mathbf{p})-f_{2}(\mathbf{p})$, where $f_{1}(\mathbf{p})$ and $f_{2}(\mathbf{p})$ are given in \eqref{f1} and \eqref{f2}, respectively given at the top of the next page. Next, we upper bound the function, $f_{2}(\mathbf{p})$, with its first-order Taylor series expansion making overall function lower bound on the sum SE i.e.,
\begin{equation}
\begin{aligned}
\sum_{k=1}^{K_s}{\rm SE}_k(\mathbf{p}) &= f_1(\mathbf{p}) - f_2(\mathbf{p}),\\
&\geq f_1(\mathbf{p}) - \bar{f}_2(\mathbf{p},\mathbf{p}_0),
\end{aligned}
\end{equation}
where $\bar{f}_{2}(\mathbf{p},\mathbf{p}_0)$ is first order Taylor's series expansion of $f_2(\mathbf{p})$ around $\mathbf{p}_0 = [p_{011},\ldots,p_{01L},\ldots,p_{0K1},\ldots,p_{0KL}]^T$ given by
\begin{equation}
\bar{f}_{2}(\mathbf{p},\mathbf{p}_0) = f_2(\mathbf{p}_0) + \nabla f_2(\mathbf{p}_0)^T(\mathbf{p} - \mathbf{p}_{0}).
\end{equation}
\begin{figure*}[t!]
	\begin{align}
	f_{1}(\mathbf{p}) &=  \sum_{k=1}^{\numSUEs}{\rm log}_2\left(\varsigma_k^2 + \sum_{i=1}^{\numSUEs}\sum_{l=1}^\numSAPs p_{kl}b_{ikl}+ \sum_{i=1,i\neq k}^{\numSUEs}\sum_{l=1}^{L}\sum_{l^{'}=1}^{L}\sqrt{p_{il}}\sqrt{p_{il^{'}}}a_{ikl}a_{ikl^{'}}+ \sum_{l=1}^{\numSAPs}\sum_{l^{'}=1}^{\numSAPs}\sqrt{p_{kl}}\sqrt{p_{kl^{'}}}a_{kkl}a_{kkl^{'}}\right)\label{f1}\\
	f_{2}(\mathbf{p})&= \sum_{k=1}^{\numSUEs} {\rm log}_2\left(\varsigma_k^2 + \sum_{i=1}^{\numSUEs}\sum_{l=1}^\numSAPs p_{kl}b_{ikl}+ \sum_{i=1,i\neq k}^{\numSUEs}\sum_{l=1}^{L}\sum_{l^{'}=1}^{L}\sqrt{p_{il}}\sqrt{p_{il^{'}}}a_{ikl}a_{ikl^{'}}\right) \label{f2}
	\end{align}
	\hrulefill
\end{figure*}
With the above approximation the equivalent EE maximization problem in \eqref{prblm2} transforms to a fractional concave-linear problem, which can be solved efficiently using Dinkelbach's algorithm \cite{zappone2015energy} to obtain the global optimal solution. The convex sub-problem that is needed to be solved in iteration $n$ is given by
\begin{equation} \label{DinkSubPrblm}
\mathcal{P}_n :\quad  \begin{array}{ll}
\underset{\mathbf{p}}{\rm{max.}} & F(\lambda_n) \\
\text{s.t.} & \eqref{conPwrAPs}, \eqref{conInt},\text{ and } \eqref{conPstvPwr},
\end{array} 
\end{equation}
where $ F(\lambda_n) = f_1(\mathbf{p}) - \bar{f}_2(\mathbf{p},\mathbf{p}_0) - \lambda_n \overline{P}_T(\mathbf{p})$. Let ${\rm EE}_{\rm lb}(\mathbf{p},\mathbf{p}_0) = (f_1(\mathbf{p}) - \bar{f}_2(\mathbf{p},\mathbf{p}_0))/{\overline{P}_{T}(\mathbf{p})}$ denote the lower bound on the objective function \eqref{eeObj2}.
Algorithm \ref{Algo}, presents the pseudo-code to obtain the solution for \eqref{prblm2} with ${\rm EE}_{\rm lb}(\mathbf{p},\mathbf{p}_0)$ as the objective. The computational complexity of the sub-problem in \eqref{DinkSubPrblm} is polynomial in the number of variables and constraints\cite{zappone2015energy}.

It is worth noting that the computation of the power coefficients is done at the CPU and is dependent only on statistics of all channel links. 
Since, channel statistics vary slowly, the power coefficients computed remain valid for a long time. 
On the other hand S-APs compute the precoders locally based on the instantaneous  channel estimates. 
Hence, only power control coefficients need to sent from CPU to the APs, which reduces the data exchange over the front-haul links.  
\begin{algorithm}\small
	\caption{Sequential Dinkelbach's Algorithm}
	\begin{algorithmic} \label{Algo}
		\STATE  1. For a given $P_{\rm max};\ \mathcal{I}_{\rm th};$ Choose $\mathbf{p}_0$ from feasibility set; 
		\STATE \quad Initialize $ t = 0$;
		\STATE 2. {\bf Repeat }
		\STATE \qquad a. Initialize $\epsilon > 0;\ n = 0;\lambda_n = 0;$
		\STATE \qquad b. {\bf Repeat}
		\STATE \qquad \qquad (i). n = n+1;
		\STATE \qquad \qquad (ii). Solve the sub-problem $\mathcal{P}_n$ in \eqref{DinkSubPrblm}; Let the optimal 
		\STATE \qquad \qquad \qquad power coefficients vector be $\mathbf{p}^{*}_n$;
		\STATE \qquad \qquad (iii). $F(\lambda_n) =f_{1}(\mathbf{p}^{*}_n) - \bar{f}_{2}(\mathbf{p}^{*}_n,\mathbf{p}_t) - \lambda_n \overline{P}_T(\mathbf{p}^{*}_n)$;
		\STATE \qquad \qquad (iv). $\lambda_n = \dfrac{f_{1}(\mathbf{p}^{*}_n) - \bar{f}_{2}(\mathbf{p}^{*}_n,\mathbf{p}_t)}{\overline{P}_T(\mathbf{p}^{*}_n)}$;
		\STATE \qquad \quad ~{\bf Until} $F(\lambda_n)\leq \epsilon$
		\STATE \qquad c. Set $t = t+1$;
		\STATE \qquad d. Set $\mathbf{p}_t = \mathbf{p}^{*}_n$;
		\STATE  \quad ~{\bf Until} ${\rm EE}_{\rm lb}$ converges.
	\end{algorithmic}
\end{algorithm}

\section{Numerical Results}
\begin{figure}[!htbp]
	\centering
	\includegraphics[width=0.65\linewidth]{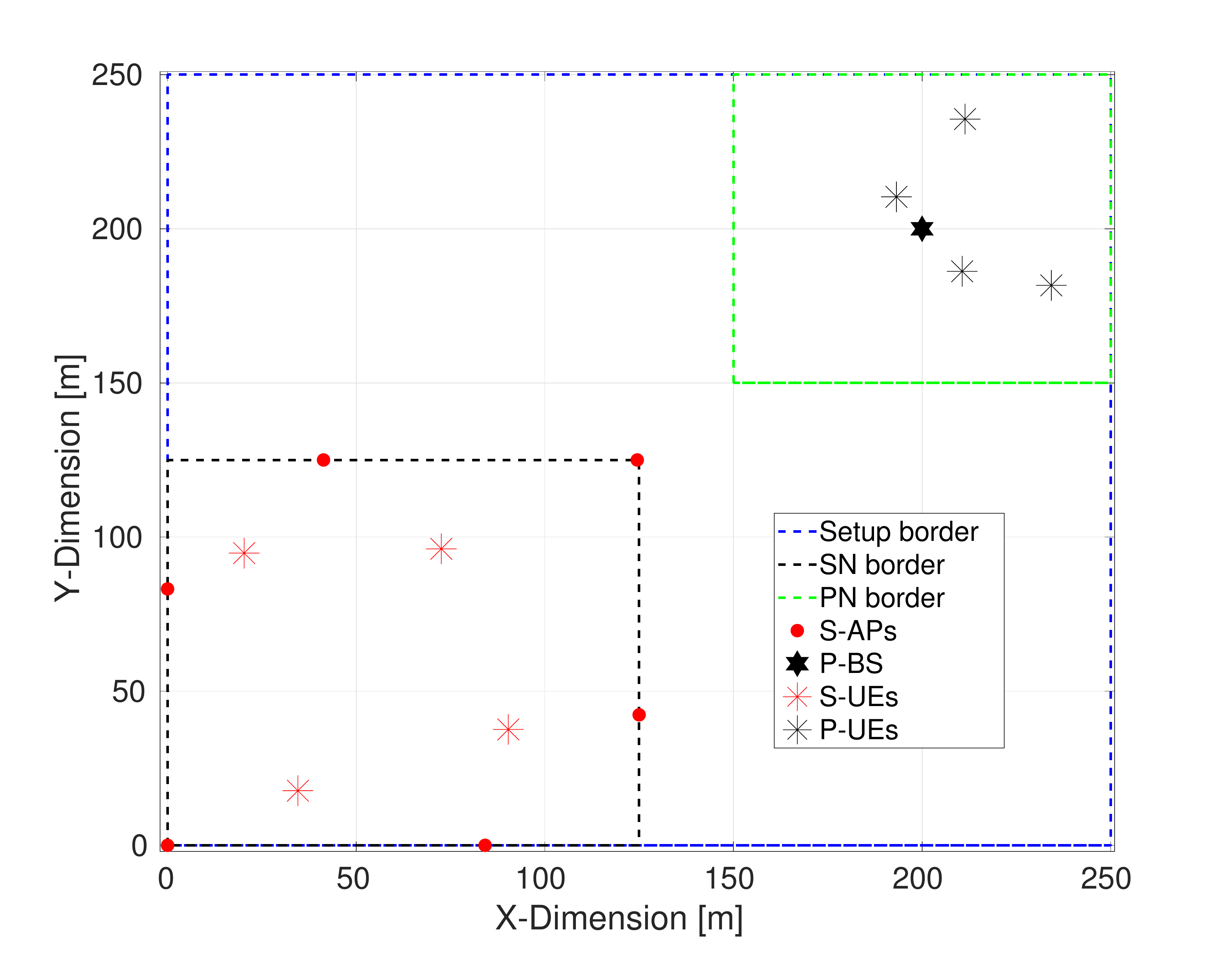}
	\caption{An example of the simulation setup considered.}
	\label{fig:SimSetup}\vspace{-2mm}
\end{figure}

We now evaluate the performance of the proposed power allocation algorithm using numerical simulations.
Our simulation setup is as follows. The secondary RadioWeave network is in a room of size $125~{\rm m}\times 125~{\rm m}$ and PN is an outdoor setup with users distributed in an area of  $100~{\rm m}\times 100~{\rm m}$. The S-APs are placed on the walls of the room with equal spacing and at a  vertical height of $5~{\rm m}$ above the floor. The S-UEs are uniformly distributed inside the room.
Similarly, P-UEs are uniformly distributed in the PN area and its base station located in the center of PN with vertical height of  $5~{\rm m}$ above the plane of S-UEs/P-UEs. We have set $\numSAPs = 6, \numAntennasSAP = 4, \numSUEs=4,\numAntennasPBS = 5$, and  $\numPUEs = 4$. One instant of the simulation setup is shown in Fig. \ref{fig:SimSetup}.

\begin{figure}[!htbp] \vspace{-2mm}
	\centering
	\includegraphics[width=0.5\textwidth]{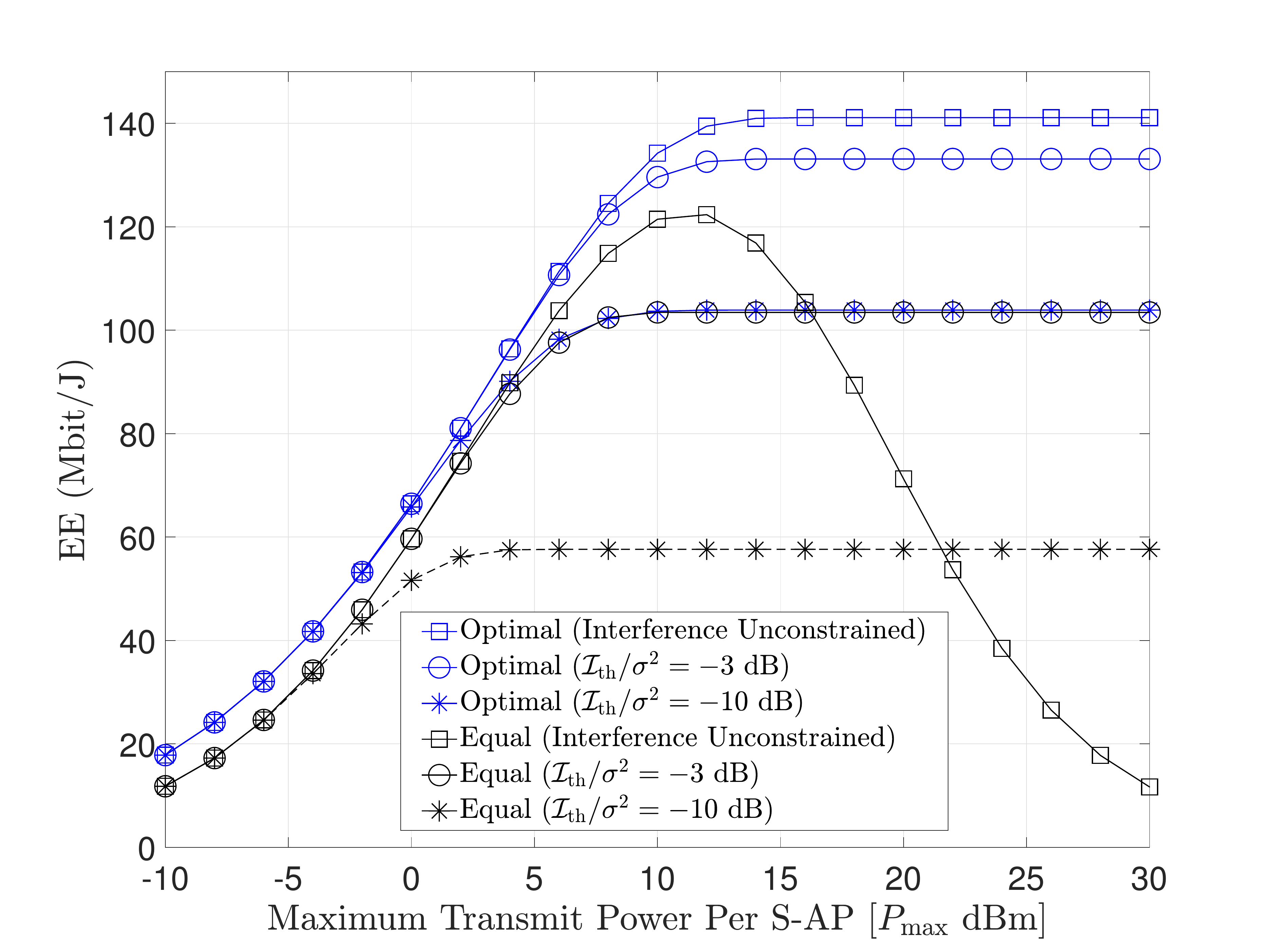}
	\caption{{Comparison of optimal and equal power allocation schemes with EE versus $P_{\rm max}$  at S-APs under interference and power constrained regimes.}} \vspace{-8 mm}
	\label{FigEEvsPwr}
\end{figure}

Let ${\rm d}_{kl}$ denote the distance between S-UE $k$ and S-AP $l$.
The thermal noise variance and path-loss between S-APs and S-UEs are modeled as
$\sigma^2 = -174 + 10\log_{10}({\rm B}) + {\rm NF}$~dBm and
$\beta_{kl} = -30.5 + 36.7\log_{10}({\rm d}_{kl}/{1~{\rm m}})$~dB, respectively,
where $B = 20~{\rm MHz}$, and the noise figure ${\rm NF} = 9~{\rm dB}$.
The same path loss model is considered between P-UEs and S-APs. We consider $\tauc = 2000$, $\taup = 8$, $\xi = 0.25 {\rm W/(Gbit/s)}, \zeta = 1.4,\ {\rm PC} = 1 {\rm W}$. We assume that P-BS employs MR precoding (normalized similar to \eqref{mrPrecod}) with its MMSE channel estimates $\{\ppchannelest{i},\ i=1,\ldots,\numPUEs\}$ and accordingly $\mathbf{Q}_{\rm p}$ can be computed as
 \begin{equation}
 \mathbf{Q}_p = \sum_{i=1}^{\numPUEs} \frac{\widehat{\mathbf{D}}_i}{\trace{\widehat{\mathbf{D}}_i}},
 \end{equation}
 where $\widehat{\mathbf{D}}_i = \expect{\ppchannelest{i}\ppchannelest{i}^H}$.
 
 The spatial correlation is modeled using a local scattering model where each AP has a uniform linear array with half wavelength antenna spacing and the multipath components are Gaussian distributed in the angular domain with a 15 degree standard deviation around the nominal angle to the user. We solved each iteration of Dinkelbach's algorithm using an interior-point method. All the plots are average over 100 setups of uniformly distributed S-UEs/P-UEs.

We compare the performance of the proposed power allocation algorithm with a simpler equal power allocation scheme, in which 
\begin{equation}
	p_{il} = \min \left\{\frac{\Pmax}{K_{\rm s}},\frac{\Ith}{V_1},\ldots,\frac{\Ith}{V_{K_{\rm p}}}\right\},
\end{equation}
where $V_m = \sum_{l=1}^{\numSAPs}\sum_{i=1}^{\numSUEs}\vartheta_{iml}$.

Fig.~\ref{FigEEvsPwr} plots  EE as a function of the maximum transmit power $P_{\rm max}$ for different values of the interference threshold normalized by the noise power i.e., $\mathcal{I}_{\rm th}/\sigma^2$. 
The performance of both optimal power allocation and equal power allocation corresponding to interference constrained and interference unconstrained regimes are shown. For small values of $\Pmax$, the EE increases with increasing $\Pmax$. The interference constraint in this region is inactive and the performance is limited by $\Pmax$. As $\Pmax$ increases the EE saturates to a value dependent on $\Ithn$. 
{\em Interference Constrained Regime} ($\Ithn=-3$~dB and $\Ithn=-10$~dB): In this regime, the EE depends on the value of $\Ithn$. This happens at $\Pmax=14$~dBm for $\Ithn=-3$~dB and  $\Pmax=10$~dBm for $\Ithn=-10$~dB. Here, the interference constraint is active and the performance is independent of $\Pmax$.
We see a similar trend for equal power allocation scheme as well. 
However, the proposed power allocation performs significantly better. For example, when $\Pmax=15$~dBm and $\mathcal{I}_{\rm th}/\sigma^2 = - 3{\rm dB}$  its EE 30 ${\rm Mbit/J}$ higher than the equal power allocation. 

{\em Interference Unconstrained Regime}:
In this regime, the EE of optimal power allocation increases for small $\Pmax$ and saturates for large $\Pmax$. This is because the optimal power allocation algorithm transmits with a total power lower than $\Pmax$ to maximize the EE. The saturation happens at $\Pmax=14$~{\rm dB}. However, the trend for the equal power allocation is not similar. It increases as $\Pmax$ increases initially and then decreases. This is because at high power the sum throughput grows logarithmically while the total power grows linearly causing EE to decline rapidly.

\begin{figure}[!htbp] 
	\centering
	\includegraphics[width=0.45\textwidth]{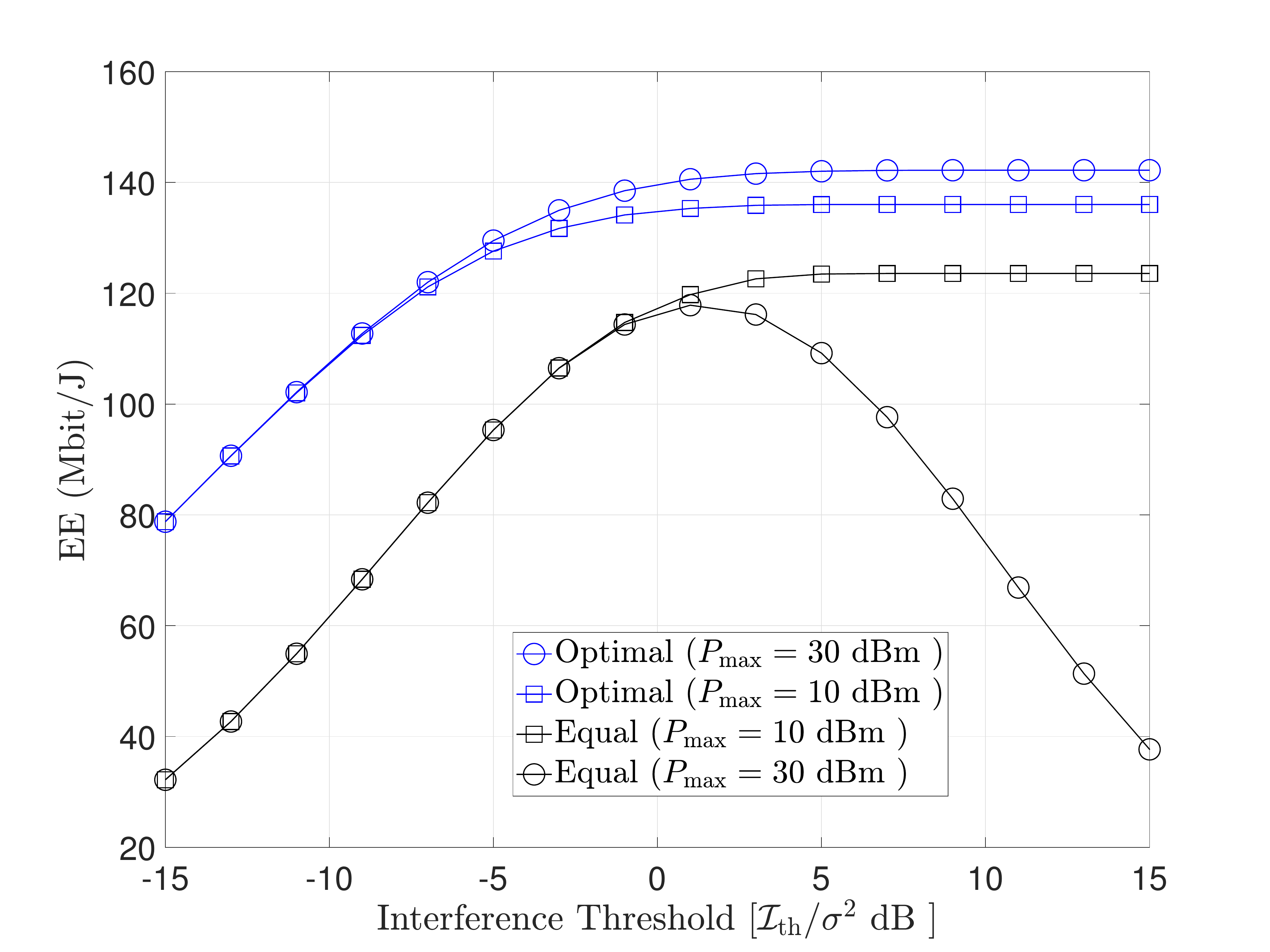}
	\caption{{Comparison of optimal and equal power allocation schemes with EE versus $\mathcal{I}_{\rm th}$ at P-UEs under interference and power constrained regimes.}}
	\label{FigEEvsInt}
\end{figure}

Fig.~\ref{FigEEvsInt} plots the EE as a function of the normalized interference threshold i.e., $\Ithn$. Similar to Fig.~\ref{FigEEvsPwr} the performance for the optimal and equal power allocation policies are shown. {\em Interference Constrained Regime}: In this regime, EE increases with increase in interference threshold, invariant to $\Pmax$. This is observed in Fig.~\ref{FigEEvsInt} up to $\Ithn = -8~{\rm dB}$. A similar trend is observed for equal power allocation up to $\Ithn = -2~{\rm dB}$. {\em Power Constrained Regime}: In this regime, EE is limited by $\Pmax$ and the interference constraint is inactive. It can be observed at $\Ithn = 4~{\rm dB}$ for both $P_{\rm max}=30~{\rm dBm}$ and $P_{\rm max}=10~{\rm dBm}$. A similar trend is observed for equal power allocation with $P_{\rm max}=10~{\rm dBm}$. However, with $P_{\rm max}=30~{\rm dBm}$, equal power allocation has a different behavior. This trend is similar to the reasons as explained for Fig.~\ref{FigEEvsPwr} i.e., for the interference unconstrained regime more power (linear increase) is allocated to gain (logarithmic gain) very little sum throughput causing the EE to decline as shown in Fig.~\ref{FigEEvsInt}.
\vspace{-5mm}

\section{Conclusion}
In this paper, we proposed a power allocation policy that maximizes the EE of an average interference constrained secondary RadioWeave network.
The iterative  algorithm proposed requires only channel statistics and enables efficient utilization of energy and spectrum.
Our numerical results demonstrated that the behavior of EE depends on different regime of operation.
They show that the EE limited either by the interference constrained regime or the power constrained regime.
Furthermore, they show that the proposed algorithm performs significantly better than the simple equal power allocation.

\bibliographystyle{IEEEtran}
\bibliography{reff}

\end{document}